\documentclass[
10pt,twocolumn,notitlepage,oneside]{revtex4}
\usepackage{rotating}
\usepackage{amsmath}
\usepackage{amsfonts,longtable}

\begin{document}

\title{Early-Type Disk Galaxies:  Structure and Kinematics}

\author{\firstname{A.~V.}~\surname{Zasov}}
\affiliation{Sternberg Astronomical Institute,  Moscow State
University,
 \protect\\
 Universitetski{\u\i} pr. 13, Moscow, 119899 Russia}

\author{\firstname{A.~V.}~\surname{Moiseev}}
\affiliation{Special Astrophysical Observatory, Russian Academy
of Sciences,
 \protect\\
 Nizhni{\u\i} Arkhyz, 369167 Karacha{\u\i}-Cherkessian
Republic, Russia}

\author{\firstname{A.~V.}~\surname{Khoperskov}}
\affiliation{Volgograd State University, Volgograd, 400068 Russia}

\author{\firstname{E.~A.}~\surname{Sidorova}}
\affiliation{Volgograd State University, Volgograd, 400068 Russia}

\begin{abstract}
Spectroscopic observations of three lenticular (S0) galaxies
(NGC~1167, NGC~4150, and NGC~ 6340) and one SBa galaxy (NGC~2273)
have been taken with the 6-m telescope of the Special Astrophysical
Observatory of the Russian Academy of Sciences aimed to study the
structure and kinematic properties of early-type disk galaxies. The
radial profiles of the stellar radial velocities and the velocity
dispersion are measured. $N$-body simulations are used to construct
dynamical models of galaxies containing a stellar disk, bulge, and
halo.  The masses of individual components are estimated for
maximum-mass disk models. A comparison of models with estimated
rotational velocities and the stellar velocity dispersion suggests
that the stellar disks in lenticular galaxies are ``overheated'';
i.e., there is a significant excess velocity dispersion over the
minimum level required to maintain the stability of the disk. This
supports the hypothesis that the stellar disks of S0 galaxies were
subject to strong gravitational perturbations. The relative
thickness of the stellar disks in the S0 galaxies we consider
substantially exceed the typical disk thickness of spiral galaxies.
\end{abstract}

\maketitle

\section{INTRODUCTION}

Early-type disk galaxies are galaxies of morphological types
S0\mbox{--}S0/a (lenticular galaxies) and Sa with properties similar
to those of lenticular galaxies. The structure of early-type disk
galaxies is similar to that of later-type spirals: they have a
massive stellar disk and, in many cases, also a dynamically
decoupled stellar circumnuclear disk, a developed bulge, and a dark
halo that determines the rotational velocity of their outer regions.
They differ from most later-type spirals in their higher (on
average) bulge luminosities, the low contrast or even total lack of
their spiral arms, the very low surface density of gas ({HI}), and,
consequently, their extremely weak star formation.

Explaining the observed features of early-type disk galaxies poses
a number of problems. First and foremost, it is unclear whether
lenticular galaxies represent a logical extension of the  Sd--Sa
morphological sequence of galaxies, which reflects the conditions
for their formation and the nature of their ensuing ``quiescent''
evolution, or whether their peculiarities are caused by  their
interaction with the  environment (mergers, accretion of small
satellites, loss of gas due to the pressure of the external
medium).

Indeed, lenticular galaxies include many objects whose structure
suggests an appreciable external influence (e.g., dynamically and
chemically decoupled circumnuclear  disks, peculiarities of the
radial brightness profile, or peculiar structural features, such
as polar rings). S0 galaxies in rich clusters appear to form as a
result of the direct effect of the intergalactic gas on the
interstellar medium, directly or indirectly leading to a reduction
in the amount of gas and a ``halt'' of active star formation.
Environmental effects evidently play a key role in this case, as
is demonstrated by the lower percentage of early galaxies in
distant clusters (the Butcher--Oemler effect [1, 2]), and the
lower contemporary rates of the current star formation in galaxies
located in denser environments~[6]. However, early-type galaxies
also include quite a few field galaxies, which may have different
histories.

The problem of the gas content in lenticular galaxies is equally
interesting. Even when HI is present in detectable amounts, its
total mass is at least an order of magnitude lower than would be
expected to result from a simple return to the interstellar medium
of gas ejected by evolved disk stars~[4]. The scarcity of data on
the thickness of the stellar disks in   lenticular galaxies makes
it difficult to compare them with other galaxies in terms of the
volume gas density or the gas pressure in the disk plane.

The relative mass fraction of  the dark halos in early disk
galaxies also remains an open question. This is due, first and
foremost, to difficulties in estimating rotation curves at large
galactocentric distances based on stellar absorption lines. Fairly
extended HI rotation curves have been obtained for a small number
of lenticular galaxies. According to Noordermeer~[5] and
Noordermeer et al.~[6], the rotational velocities in these
galaxies often decrease toward the periphery, but they still imply
the presence of fairly massive dark halos.

The large scatter of data in the Tully\mbox{--}Fisher relation
(luminosity--rotation velocity diagram) for lenticular galaxies
suggests substantial inhomogeneity of their properties (see,
e.g.,~[7] and references therein). It also follows from the analysis
of the velocity dispersions of old stars in the galaxy disks, which
shows that in some early-type disk galaxies, the stellar velocity
dispersion substantially exceeds the minimum level required for the
gravitational stability of the disk, whereas, in later-type spiral
galaxies, the velocity dispersion of the disk stars usually appears
to be close to its threshold value~[8]. However, the relatively low
accuracy of the estimated velocity dispersions for disk stars beyond
the bright bulges and problems with decomposing the velocity
dispersion into $r, \varphi$, and $z$ components lead us to treat
this conclusion as being tentative.

A comparison of the observational data with dynamical models in
which the velocity dispersion of the disk stars---both in the plane
of the disk and perpendicular to this plane---is close to the
critical values for the dynamical stability of the disk provides
insight into its dynamical evolution. A comparison of model
($c_{\ell}$) and observed ($c_{\textrm{obs}}$) line-of-sight
velocity dispersions for old disk stars can reveal one of three
possible situations.

(1) $c_{\textrm{obs}}<c_{\ell}$. If interpreted in terms of disk
stability, it would imply that the mass of the disk adopted in the
model is overestimated, and that the disk must be ``less massive''
in order to satisfy the conditions of dynamical stability against
perturbations in the plane of the disk and against bending
perturbations.

(2) $c_{\textrm{obs}} = c_{\ell}$ within the measurement  errors. In
this case, it appears that dynamical instabilities during the
formation of the bulk of the disk mass have brought the stellar disk
to a marginally stable state, where the stellar velocity dispersion
is determined by the surface density of the disk, its rotational
velocity as a function of galactocentric distance $r$, and its
internal structure. This appears to be the most common case among
noninteracting spiral galaxies.

(3) $c_{\textrm{obs}}> c_{\ell}$; i.e., the observed stellar
velocity dispersion exceeds the model values corresponding to
marginal disk stability. In this case, there is reason to believe
that disk stars have acquired an excess (i.e., above the level
required for stability of the disk) energy of random motions
during their evolution, so that the disk has become overheated.
This can be viewed as evidence that the stellar population of the
disk subsystem of the galaxy has been subject to strong
gravitational perturbations, for example, as a result of mergers
of massive stellar or gaseous satellites, or of close interactions
with nearby neighbors. In principle,  the dynamical heating in the
inner part of the galaxy can also be related to the disruption of
a high-contrast stellar bar.

\begin{figure*}
%%% Figure:1
\includegraphics{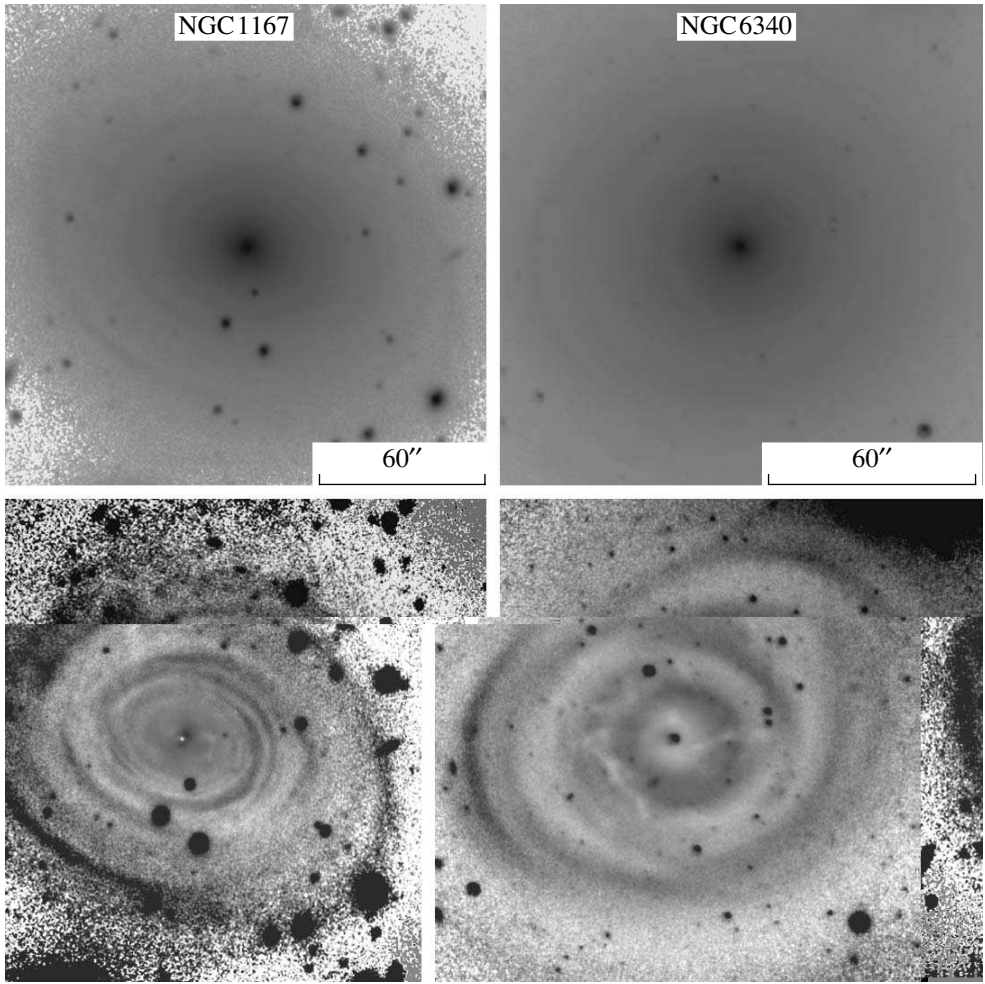}
\caption{$V$-band images of the galaxies NGC~1167 (left) and
NGC~6340 (right) on logarithmic brightness scales, showing the raw
images (top) and the images after the  model surface brightness
distribution subtracted (bottom).}
\end{figure*}

Constructing dynamical galaxy models that can yield estimates of
the disk-to-halo mass ratio or disk thickness requires estimates
of the rotation velocity and stellar velocity dispersion at the
largest possible galactocentric distances, preferably along the
main  axes of the galaxy, to make it possible to determine the
velocity dispersion along both the radial and vertical directions.

In this paper, we describe spectroscopic observations and the
results of our construction of dynamical models for four early-type
disk galaxies. Table~1 lists the basic parameters of these galaxies
we have adopted. Figure~1 shows images of two of the galaxies
with low-contrast structure taken with the 6-m telescope of the
Special Astrophysical Observatory of the Russian Academy of
Sciences. Table~2 gives a log of the observations.

The galaxy luminosities in Table~1 correspond   to the total $B_T$
magnitudes from the   \mbox{HYPERLEDA} database [11]. The last
column gives the radius of the exponential stellar disk
$R_{\textrm{max}}$ which was adopted in galaxy models. This radius
corresponds to the isophotal radius $R_{25}=D_{25}/2$, or the
radius beyond which the photometric profile steepens.

\section{DESCRIPTION OF INDIVIDUAL GALAXIES}

\textbf{NGC~1167.} This galaxy contains a substantial amount of
{HI}. According to the observations of Noordermeer et al.~[14], the
total {HI} mass is $1.7\times 10^{10}$~$M_{\odot}$, but the gas is
distributed over a very large area, so that the average surface
density $\langle {\textrm{HI}}\rangle$  within the optical radius
$R_{25}$ is less than $2 $~$M_{\odot}/$pc$^2$. The gas density
remains below the critical value required for gravitational
instability of the gaseous layer at all galactocentric distances
$r$, explaining the lack of observed star-forming regions in the
galaxy. HI observations show an extended rotation curve, which
slowly decreases after reaching its maximum and extends over 10
radial disk scales [5,~6]. The maximum rotational velocity of the
disk is almost 400~km/s, making this one of the most rapidly
rotating and massive disk galaxies known.

The total bulge luminosity of this galaxy is about one-third of
the disk luminosity~[15]. The bulge has a steep photometric
profile (our $V$-band measurements yield a Sersic parameter of
$n\simeq 3$) so that its brightness dominates the central region,
which has a size of several kpc. Starting from $r= 15''{-}20''$
($5.0{-}6.5$~kpc), the photometric profile becomes exponential,
and hence the bulge's contribution to the observed brightness
becomes small.

\begin{table*}[t!]
%%% Table:1

\caption{Adopted parameters of the four galaxies}
%\settabpars{2.5}{3}{1.5}
%\newcolumntype{a}{D{.}{.}{3.1}}
%\newcolumntype{b}{D{.}{.}{2.2}}
%\newcolumntype{d}{D{.}{.}{1.6}}
%\stretchtab[1.005]{%
\begin{tabular}{ccccccc}
 \hline
   \multicolumn{1}{c}{Galaxy} & \multicolumn{1}{c}{Type} &
   \multicolumn{1}{c}{Tilt}& \multicolumn{1}{c}{Distance,} & \multicolumn{1}{c}{Luminosity,}  & \multicolumn{1}{c}{Radial scale} & \multicolumn{1}{c}{$R_{\textrm{max}}$,} \\[-5pt]
             &     & \multicolumn{1}{c}{angle $i$}    &  \multicolumn{1}{c}{Mpc }    &  \multicolumn{1}{c}{$10^{10}L_{\odot}$} & \multicolumn{1}{c}{of the disk $r_d$,~kpc}&   \multicolumn{1}{c}{kpc}   \\
  \hline
  \hline
  NGC~1167 & S0 & ~~~36$^{\circ}$ [6] & 67  & 10 & 8.0 [6] & 31.9 \\
  NGC~2273 & SBa& ~~~50$^{\circ}$ [9]& 25.7& 1.48 & 3.7 [10]& 15\\
  NGC~4150 & S0 & ~~~56$^{\circ}$ [11]  & 14  & 0.33 &0.84 [12]& 3.34 \\
  NGC~6340 & S0/a&~~~26$^{\circ}$ & 19.8& 1.21 & 2.4 [13] & 9.6  \\
\hline
\end{tabular}
% end \stretchtab
\end{table*}

\begin{table*}[t!]
%%% Table:2
\caption{Log of spectroscopic observations}

\begin{tabular}{cccccc}
\hline
Galaxy  & \multicolumn{1}{c}{Slit} & \multicolumn{1}{c}{Date}  & \multicolumn{1}{c}{$T_{\textrm{exp}}$, s}  & Seeing& \multicolumn{1}{c}{PA}  \\[-5pt]
           & \multicolumn{1}{c}{direction}&                   &           &&  \\
\hline \hline
NGC~1167   & Major axis &25/26.10.2005& 8400         &   3$^{\prime\prime}$   & 70$^{\circ}$\\
           & Minor axis   &24/25.11.2005& 9600         &   2.2          & 160\\
\hline
NGC~2273   & Major axis &26/27.11.2005& 8400         &   2.5          & 58\\
           & Minor axis   &25/26.12.2005& 4800         &   2.6          & 148\\
\hline
NGC~4150   & Major axis &02/03.02.2005& 9600         &   2.8          & 146\\
           & Minor axis   &03/04.02.2005& 7200         &   1.4          & 57\\
\hline
NGC~6340   & Major axis &06/07.05.2005& 7200         &   3      & 120\,\\
           & Minor axis   &02/03.06.2006& 9000         &   2.1          & 30\\
\hline
\end{tabular}
% end \stretchtab
\end{table*}

The images taken with the 6-m telescope of the Special
Astrophysical Observatory reveal a system of ring-like arcs or
spirals, which were also noted by Noordermeer~[5]. The spirals are
remarkably thin and smooth, without characteristic irregularities
and bright knots (usually due to local star-forming regions in the
galaxy's spiral arms), suggesting a lack of O stars in these
structures. The nature of the thin, smooth arms remains unclear.
According to preliminary estimates we obtained by analyzing images
of the galaxy, the $B{-}V$ color index is slightly
($0.05^m-0.1^m$) bluer in the spirals than in the surrounding
disk, suggesting ongoing or recent star formation. If the initial
stellar mass function in these spirals is  anomalously steep
slope, this could explain the lack of massive stars capable of
producing extended, bright HII regions.

There are no other galaxies of comparable luminosity in the
neighborhood of NGC~1167.

\textbf{NGC~2273.} This is a relatively isolated galaxy of moderate
luminosity with a Seyfert nucleus and an unusually clear and
symmetric structure in the central region, with a radius of
${\approx} 25''$ (a bright bar and a pseudo-ring) and faint fuzzy
spirals beyond it. The photometric profile of the galaxy cannot be
described by a simple exponential law: it steepens at $r > 80'' {-}
100''$~[10, 15]. Like NGC~1167, NGC~2273 contains a fairly large
amount of gas~[6, 16]. At $r<40''$, the brightness profile is
dominated by the bulge. The total hydrogen mass is $M_{\textrm{HI}}
= 2.42\times 10^9 $~$M_{\odot}$, but the average gas surface density
is roughly as low as it is in NGC~1167 ($\langle
{\textrm{HI}}\rangle = 2.2$~$M_{\odot}/$pc$^2$)~[6]. Moiseev et
al.~[9] and the SAURON team~[17] analyzed the optical spectra of the
galaxy in order to study the two-dimensional velocity distribution
of gas and stars. In the bar region ($r<30''$), gas (but not the
stars!) exhibits appreciable circular motions~[9]. The rotation
curve of the galaxy appears to have a local maximum near the center
($r\simeq 10''$), but Noordermeer et al.~[6] believe that this may
be due to noncircular gas motions. A small circumnuclear disk
coincident with the molecular-gas disk~[9] can be seen at the center
of the galaxy, visible in emission lines.

\textbf{NGC~4150.} This is a low-luminosity S0 galaxy with a smooth
photometric profile, which is fit well by an exponential law from
the very center out to at least $r = 80''$~[12]. The central region
shows traces of dust observed against the bright stellar background.
UV observations reveal a bright nucleus, indicative of the presence
of young stars~[18]. The inner part of the galaxy exhibits small
amounts of molecular and atomic gas~[4]. Distance estimates for this
galaxy are uncertain. Karachentsev et al.~[19] suggested that
NGC~4150 is located at the periphery of the Virgo cluster at a
distance of about 20~Mpc. Sage and Welch~[4] adopted a distance of
9.7~Mpc.   We adopt here a distance of 14~Mpc. The kinematics of the
inner region of the galaxy was studied earlier as part of the SAURON
project~[20]. The circumnuclear region ($r<5''$) appears to host a
counter-rotating disk.

\textbf{NGC~6340.} It is an S0/a galaxy,   a member of a group,
although there are no other galaxies of comparable luminosity in
its immediate vicinity. The image of the galaxy features a bright
inner disk---lens---and low-contrast, outer spiral arms. Long,
conspicuous dust lanes of unusual shape can be seen in the central
part of the disk, which are most likely located outside the plane
of the disk. However, the most striking features are thin
fragments of rings or spiral structures similar to those observed
in NGC~1167, but somewhat more fuzzy and confined by a narrower
zone of the disk. The galaxy has a bright bulge with a luminosity
almost half the disk luminosity. However, the bulge is fairly
strongly concentrated: its effective radii in the $I$ and $V$
bands are $r_e = 8.6''$ [13] and $r_e = 3.1''$~[21]. The
kinematics of the stellar population in the inner region of the
galaxy were analyzed by Bottema~[22] and Corsini et al.~[23].
Two-dimensional spectroscopy of the circumnuclear region of the
galaxy revealed a chemically decoupled nucleus consisting of old
stars with a relatively high abundance of heavy elements, and
probably a circumnuclear polar ring located within $r=500$~pc~[24,
25]. Following Sil'chenko~[24], we adopt for this galaxy a
distance of 19.8~Mpc.

\section{OBSERVATIONS}

We observed the galaxies with the SCORPIO multimode
instrument~[26] mounted at the primary focus of the 6-m telescope
of the Special Astrophysical Observatory. We studied stellar
kinematics using the slit spectrograph mode with a $6' \times 1''$
slit and a $2048\times2048$~pixel EEV~42--40 CCD as the detector.
The scale along the slit was $0.36''$/pixel. Observations were
made in the wavelength range $4800{-}5540$~{\AA}, which contains
numerous absorption lines of the old stellar populations in the
galaxies. The spectral resolution was 2.2~\AA, which corresponds
to an instrumental profile with $\sigma=55$~km/s in terms of the
velocity dispersion. The log of observations in Table~2 gives the
dates of the observations, the exposures ($T_{\textrm{exp}}$), the
average seeing, and the positional angle PA of the slit. We used
an IDL-based software package to reduce the data. See~[26] for a
brief description of the algorithms employed.

\begin{figure*}
%%% Figure:2
\includegraphics[scale=0.9]{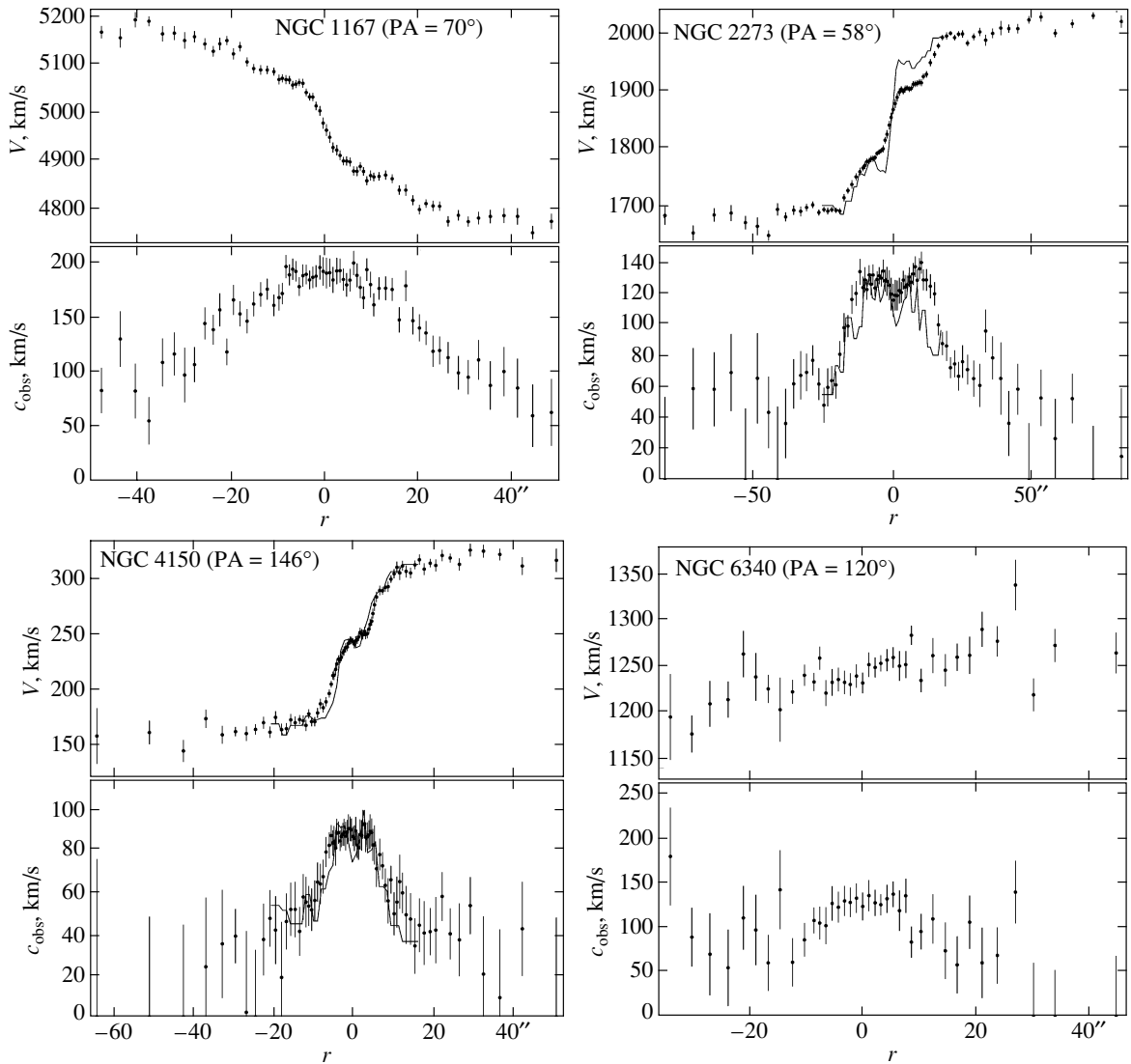}

\caption{Distribution of radial velocities and the velocity
dispersion along the major axes of the galaxies. The solid curves
show the SAURON data. }
\end{figure*}

\begin{figure*}
%%% Figure:3
\includegraphics[scale=0.9]{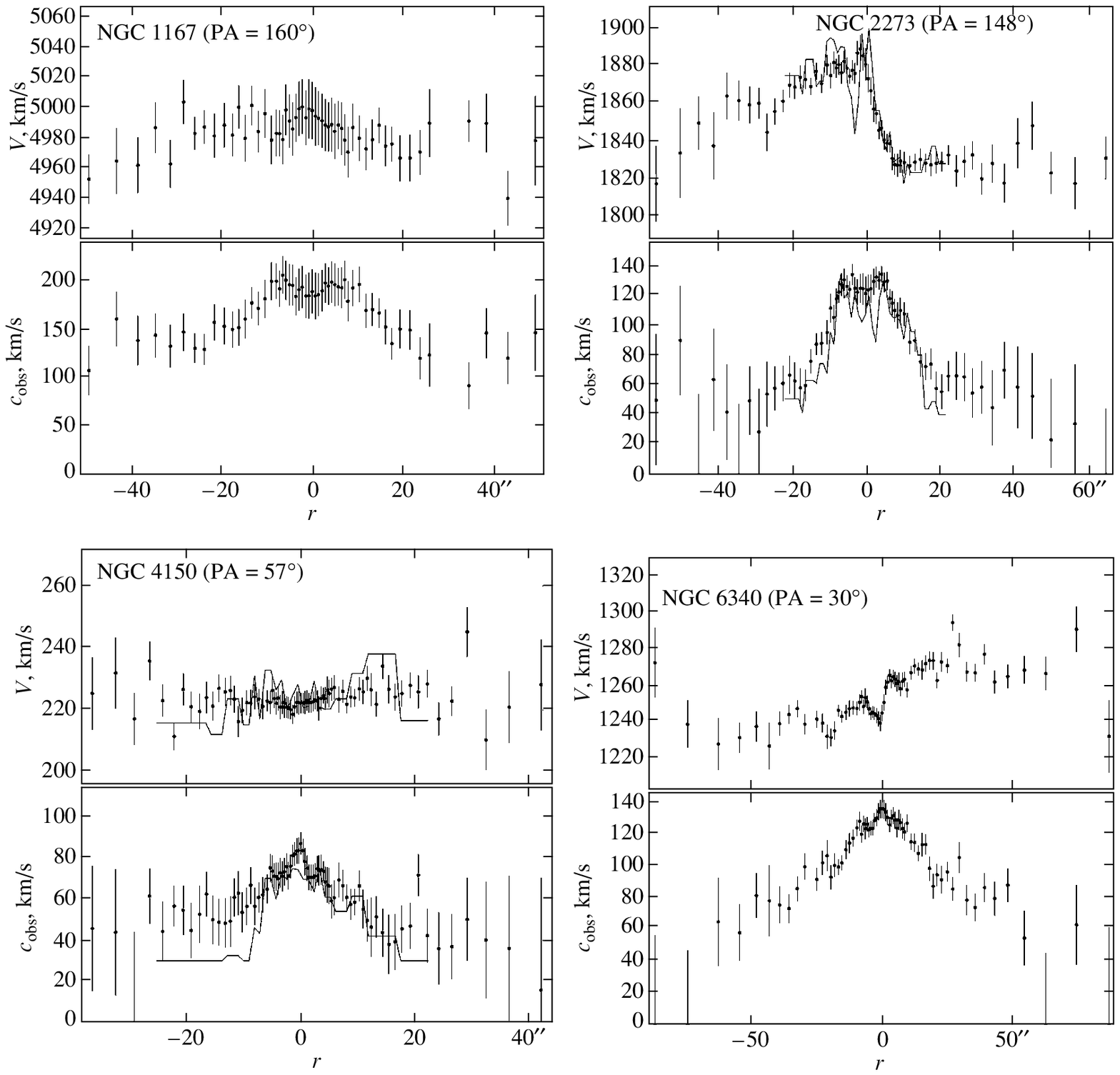}

\caption{Same as Fig.~2, but along the minor axes of the galaxies. }

\end{figure*}

We used the classical cross-correlation technique to calculate the
radial velocities and velocity dispersion of the stellar
component; the algorithms and software employed are similar to
those described by Moiseev~[27]. As templates for the cross
correlation, we used spectra of G8--K4 giants observed on the same
nights as the galaxies. To increase the signal-to-noise ratio, we
used adaptive binning (co-adding) of the spectra along the slit,
with the integration window exponentially increasing with
galactocentric distance. This technique makes it possible to compensate
for the radial exponential decrease in the surface brightness in the galaxy
disks. Figures~2 and 3 show the results of our measurements of
the variations of the radial velocities and the velocity
dispersions along the major and minor axes of the galaxies. For
comparison, these figures also show the distributions of these
parameters for NGC~2273 and NGC~4150
obtained by taking appropriate sections across the
two-dimensional SAURON velocity and velocity-dispersion maps~[17,
20]. It is obvious from this comparison that any
systematic bias of our velocity-dispersion estimates does not
exceed the measurement errors. The difference between the radial-velocity
curves along the major axis of NGC~2273 is apparently
due to the better spatial resolution of the SAURON data.

Note that some of our observations were made under unfavorable
photometric conditions (some cloudiness), which prevented us from
taking spectra of the outer regions of NGC~6340 along the major
axis. However, the high accuracy of the measurements along the
minor axis enabled us to obtain a detailed radial-velocity profile
for the stellar population. The non-monotonic behavior of the
velocity within ${\pm} 5''$ of the center is immediately obvious,
and is apparently due to the presence of a circumnuclear polar
ring. The radial-velocity gradient is not zero at large
galactocentric distances along the minor axis, and varies by about
40~km/s along the disk radius within $2'$ (which is equivalent to
about 6~kpc along the major axis). This behavior points toward a
noticeable misalignment of the kinematic and photometric axes of
the galaxy. According to preliminary estimates based on two
spectral sections, the PA of the kinematic minor axis of the
galaxy should be close to $0^{\circ}$. However, the low accuracy
of the measured velocity gradients along the spectrograph slit
prevented reconstruction of the detailed shape of the rotation
curve for this galaxy.

SCORPIO can be used to obtain not only spectra, but also direct
images of objects in the field of view. Short-exposure
($10{-}30$~s) $V$ images were used to aim the spectrograph slit at
the galactic nucleus to within $0.2''{-}0.3''$. We also took
deeper images shown in Fig.~1 for the two galaxies NGC~1167 and
NGC~6340, which exhibited unusual spiral structure on short
exposures. The images were obtained by subtracting from the
observed surface brightness the surface brightness in an
axisymmetric model with a smooth brightness distribution
consisting of a disk and bulge with  elliptical isophotes.

\section{GENERAL PRINCIPLES OF  MODELING}

We constructed dynamical models of the galaxies having the maximum
possible disk masses and model rotation curves consistent with the
observed rotation curves (maximum-disk models). We compared the
observations to both the rotation curve calculated for this model
and the minimum velocity dispersions for a collisionless disk that
would be sufficient to maintain its stability against radial and
bending perturbations. An analysis of the dynamical evolution of
the disks which follow from numerical simulations can be used to
obtain stable models without resort to approximate and
insufficiently trustworthy local analytical criteria. We used a
three-component model (disk, halo, and bulge) whose component
parameters yielded a circular velocity of
%\begin{gather}\label{Eq-5-circvelocity-sum-components:Zasov_n}
%V_c(r) \inRussian{=}{}\\
%\nonumber{}= \sqrt{
%(V_c^{disk}(r))^2+(V_c^{bulge}(r))^2+(V_c^{halo}(r))^2 }  ,
%\end{gather}

\begin{equation}\label{Eq-5-circvelocity-sum-components:Zasov_n}
V_c(r) = \sqrt{
(V_c^{disc}(r))^2+(V_c^{bulge}(r))^2+(V_c^{halo}(r))^2 } \,,
\end{equation}

where $V_c^{disk}(r)$, $V_c^{bulge}(r)$, and $V_c^{halo}(r)$ are the
corresponding contributions of individual components to the circular
velocity.

The  dynamical models of collisionless (stellar) disks are based
on numerical integration of the equations of motion of $N$
gravitationally interacting particles using the TREEcode program,
taking into account the external field of the ``hard'' bulge and
halo. This means that the parameters of the spheroidal subsystems
are considered to be stationary, and are described by free
parameters of the model. For NGC~4150, we also considered a model
with a ``live'' bulge, which enabled us to take into account the
bulge contribution not only to the gravitational potential, but
also to the stellar velocity dispersion. However, we found no
fundamental differences: the velocity dispersion remained
virtually unchanged beyond the effective bulge radius.

We specified the disk surface density in the form
%\begin{gather}\label{sigma(r)=exp(-r/L):Zasov_n}
%    \sigma(r) = \sigma_0   \exp(-r/r_d)
%\end{gather}

\begin{equation}\label{sigma(r)=exp(-r/L):Zasov_n}
  \sigma(r) = \sigma_0 \cdot \exp(-r/r_d)
\end{equation}
at galactocentric distances $r\leq R_{\textrm{max}}$ (Table~1).
Here, $r_d$ is the radial disk scale estimated from the brightness
distribution of the galaxy.

To reduce ambiguity in choosing a model that was consistent with
the observed rotation curve, we further assumed that the radial
scale for the disk surface-density variations in the region
covered by measurements is close to the radial scale $r_d$ of the
disk brightness  known from optical photometry (Table~1).

We started the modeling with the initial disk in an unstable
(subcritical) state with a Schwartzschild (ellipsoidal) velocity
distribution, such that the resulting collisionless disk was close
to the stability threshold. Hence, the resulting models represent
models of the maximum, marginally stable disks. The real disks may
have smaller masses than those obtained in our models without
violating the stability condition. Estimations of the masses of
stellar disks based on the condition that they be gravitationally
stable~[22, 28, 29] show that the maximum-disk model constructed
without allowance for the velocity dispersion may overestimate the
circular disk velocities by 20--30$\%$. Note, that the estimates of
the disk masses obtained using another method, namely hydrodynamical
modeling of the gas motions in the region of spiral arms, suggest
that only the inferred disk mases of slowly rotating galaxies
($V_c<150{-}200$~km/s) will be substantially lower than their
``maximum'' masses~[30].

We constructed an equilibrium dynamical model to calculate radial
profiles of the circular rotational velocities $V_c$, the rotational
velocities of the stars $V$ (particles in the model), and the
velocity dispersions $c_r$, $c_\varphi$, and $c_z$, which ensure the
marginal stability of the disk. To compare our models with the
observational data, we calculated the model line-of-sight velocity
dispersions:
\begin{multline}\label{5--Eq-c-obs(cr-cf-cz):Zasov_n}
c_{\ell}(r)  = (c_z^2 \cos^2(i)+c_\varphi^2\sin^2(i) \cos^2(\alpha)
+\\
+c_r^2\sin^2(i)\sin^2(\alpha))^{0.5},
\end{multline}
where $i$ is the inclination of the disk to the plane
of the sky and $\alpha$ is the angle between the slit direction and
the major axis projected onto the plane of the galaxy.

The technique of constructing a galaxy model whose disk is at the
limit of stability against both gravitational perturbations in the
plane of the disk and bending perturbations is described by
Khoperskov et al.~[28, 31] and Tyurina et~al. [32]. The
corresponding computations covered five to ten orbital-rotation
periods at the outer disk rim, ensuring the establishment of a
stationary state in which the velocity dispersion has virtually
stopped changing and has maintained its average value over several
rotational periods. We used an iterative algorithm with the
initial velocity dispersion successively approximating the
stability limit, as developed by Khoperskov et~al. [28]. The
iterative approach is based on a series of successive computations
involving  $N=2\times 10^5$ particles, each starting with an
initial velocity dispersion that is somewhat closer to the
critical value than in the previous case. To this end, we chose
the initial distribution of the velocity dispersions $c_r(r)$ and
$c_\varphi(r)$ to be between the initial and final values obtained
in the previous simulation.

We also performed control computations with $N=10^6$ particles to
monitor computational effects in the inferred radial
distributions of the disk parameters at the stability limit. The
results corroborated our earlier conclusion~[28] that models with
$N \gtrsim 10^5$ are adequate to determine the stability limit.

In  general,  we subdivided the construction of the model stable
equilibrium disks into the following stages.

(1) Estimating the components of the velocity dispersion along
three axes [see (3)] by analyzing the observed velocity
dispersions along the major and minor axes of the galaxy. As the
additional condition required for this task, we adopted
the Lindblad relation between the radial and  azimuthal
components of the velocity dispersion, which was tested in numerous
numerical simulations: $c_r / c_\varphi = 2\Omega / \kappa$,
where $\kappa$ is the epicyclic frequency.

(2) Determining the circular-velocity curve $V_c(r)$
from the observed rotational velocities of the stars,
$V^{obs}_\star(r)$, and the velocity dispersion $c_r(r)$. However,
when available, we adopted the gas rotation curve as the initial
circular-velocity curve.

(3) Decomposing the rotation curve into components representing
the bulge,   disk of finite thickness and halo (the maximum-disk
model).

(4) Choosing the initial conditions for the description of the
disk in the subcritical state, with the derived component
parameters used as a first approximation.

(5) Numerically computing the dynamical evolution of the disk
to obtain a model for the marginally stable disk whose
circular-velocity curve agrees with the curve obtained from the
observations.

(6) Computing the radial dependence of the line-of-sight
velocity dispersion for the resulting model with allowance
for the inclination of the disk.

(7) Comparing the model and observed line-of-sight velocity
dispersions.

(8) Computing the disk parameters: its mass,
average thickness (for the observed velocity
dispersion), and mass-to-luminosity ratio. Estimating the
fractional mass of the dark halo.

We did not compare the model velocity dispersions for disk stars
with the corresponding observed dispersions for central regions
of the galaxy, since both the form of the rotation curve and the
velocity dispersion of the disk stars are determined very uncertainly
in the bulge region.

\section{MODELS OF THE GALAXIES}

\textbf{NGC~1167.} We constructed three models for this galaxy (a,
b, c) to see the effect of the choice of parameters on the final
result. Here, n1167-{a} is the maximum-disk model; n1167-{c} can
be thought of as the minimum-disk model, which corresponds to an
$R$ mass-to-luminosity ratio of $M/L_R \approx 1.5$, which is
certainly lower than the mass-to-luminosity ratios of stellar
systems without active star formation and with normal stellar
population; and n1167-{b} has an intermediate disk mass. All three
models reproduce the rotation curve of the galaxy fairly well.

Figure~4a compares the observed rotational velocities with the
rotational velocities for the models. The n1167-{a} and n1167-{b}
models differ appreciably only in the central region, due to their
different bulge concentrations. The central disk surface densities
are almost the same in these two models: $1400$ and
$1340$~$M_{\odot}$/pc$^2$. The halo mass is somewhat higher in the
second model: the halo-to-disk mass ratios within
$R_{\textrm{max}}$ are equal to
$(M_h/M_d)_{\textrm{n1167-a}}=0.67$ and
$(M_h/M_d)_{\textrm{n1167-b}}=0.83$. In the n1167-{c} model, the
mass of the halo exceeds that of the disk.

Figure~4b shows the distributions of the observed and model
estimates of the line-of-sight velocity dispersion for the
marginal-disk model n1167-{a}. It is obvious that the
velocity dispersion in this model is substantially lower than the
corresponding observed values. The velocity dispersion becomes even
lower if we decrease the disk mass (curve shown by diamonds
in Fig.~4b). In the n1167-{c} model, the observed disk is
overheated in terms of the velocity dispersion $c_{\ell}$ by a
factor of almost three compared to the marginally stable state. A
comparison of our maximum-disk model with the observed distribution
of the velocity dispersion along the minor axis confirms that the
disk is significantly overheated.

\begin{figure}
%%% Figure:4
\includegraphics[scale=0.9]{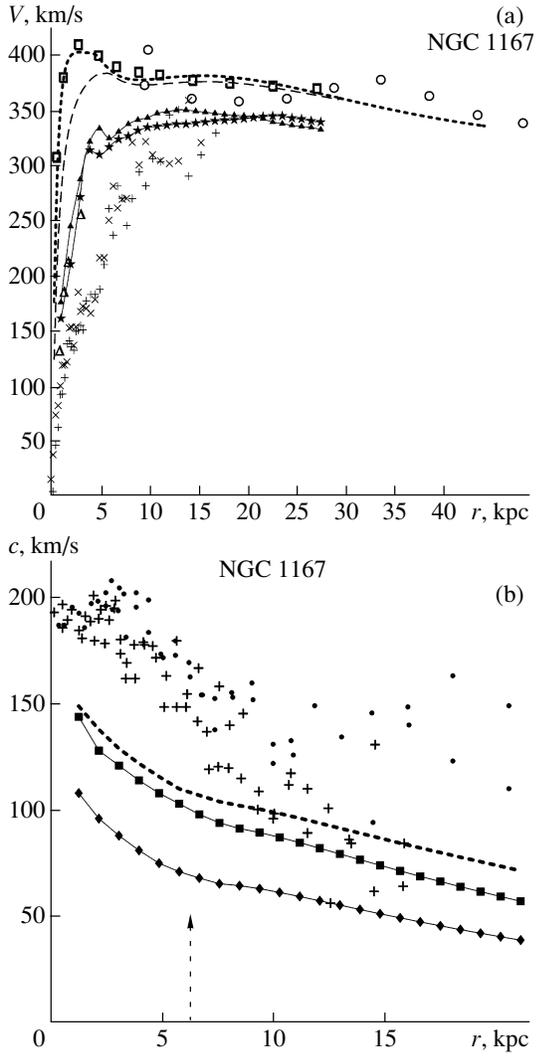}
\caption{NGC~1167. Shown are the (a) rotational velocity of the
stellar disk according to our measurements $V^{obs}_\star$ (the
straight and slanted crosses show measurements made on either side
of the center), gas rotational velocity [5] (open circles),
H$\alpha$ rotational velocities [5] (open triangles), the circular
velocity $V_{c}$ according to the data of Noordermeer et~al. [14]
(open squares), $V_c$ for the n1167\mbox{-}a model (dotted curve),
the corresponding stellar-disk rotational velocity for this model
(filled triangles), $V_c$ for the n1167-b model (dashed curve), and
the corresponding stellar-disk rotational velocity for this model
(asterisks); (b) stellar velocity dispersion $c_{\textrm{obs}}$
according to our observations along the major axis (crosses) and
minor axis (small filled circles), line-of-sight velocity dispersion
of disk stars in the maximum-disk model along the major axis (filled
squares) and minor axis (dotted curve), and the corresponding
velocity dispersion along the major axis for the n1167-c model with
a low-mass disk (diamonds). The vertical dashed arrow indicates the
conventional boundary of the bulge.}

\end{figure}

\begin{figure}
%%% Figure:5
\includegraphics[scale=0.9]{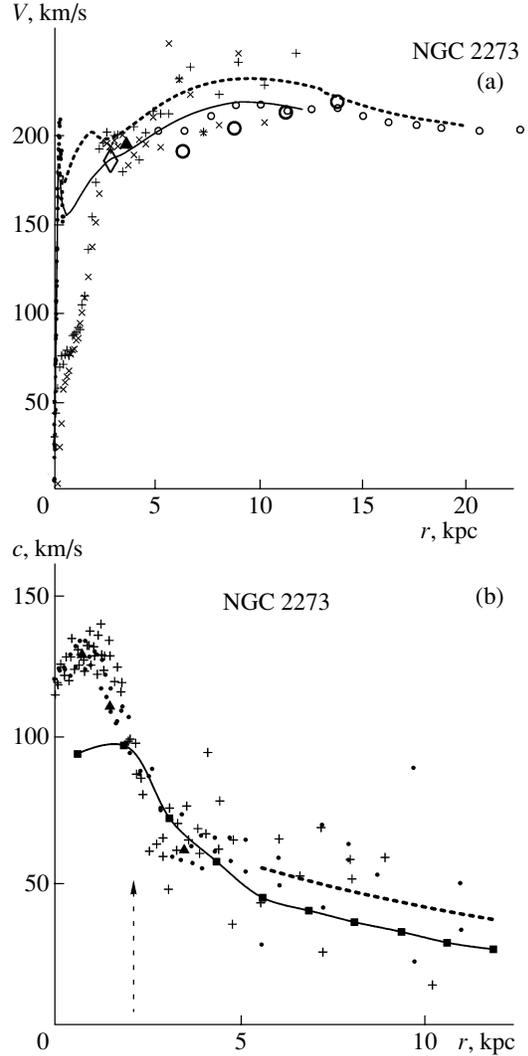}
\caption{NGC~2273. Shown are the (a) rotational velocity of the
stellar disk according to our measurements $V^{obs}_\star$ (straight
and slanted crosses show the measurements made on either side of the
center), HI rotational velocity~[5] (small open circles), rotational
velocity based on the same data calculated for a model with fixed
disk inclinations (large open circles), H$\alpha$ rotational
velocities~[5] (small filled circles near the center), the model
rotation curve (solid), the rotation curve that agrees best with the
measured rotational velocities and stellar velocity dispersions
(dotted), the SAURON-VII data for $V_{\textrm H\beta}$ (diamonds)
and $V_{\star}$ (triangle) (see text); (b) stellar velocity
dispersion according to our observations along the major axis
(crosses) and minor axis (small filled circles), the line-of-sight
velocity dispersion of the disk stars in the maximum-disk model
along the major axis (filled squares) and minor axis (dotted curve),
and the velocity dispersion according to~[17] (filled triangles).
The vertical dashed arrow indicates the conventional boundary of the
bulge.}
\end{figure}

Below we analyze only the galaxy models with the maximum disk mass.

We determined the disk half-thickness $z_0$ from the dynamical
model, with the stellar density distribution in the $z$
direction approximated by the law $\varrho\propto
\textrm{ch}^{-2}(z/z_0)$, which is valid for a self-gravitating
isothermal disk. The average disk half-thickness of this galaxy is
$z_0\approx 2.8$~kpc if calculated for the
maximum-disk model. Thus, the disk is fairly thick, both in
absolute terms and compared to its radial scale: $z_0/r_d=0.35$.
For comparison, the half-density half-thickness of the stellar
disk in the solar neighborhood is about 350~pc~[33],
which corresponds to $z_0\approx 400$~pc.

\textbf{NGC~2273.} As in the previous case, an HI rotation curve
is available for NGC~2273~[5, 16]. We used velocity-field
measurements kindly provided by E.~Noordermeer to construct the
rotation curve outside the galaxy bulge with the fixed disk
inclinations adopted in this paper (Table~1). Figure~5 shows the
rotation curves and velocity-dispersion distributions based on the
observational data. The HI rotation curve exhibits a local maximum
at a galactocentric distance of several kpc, but the model curve
reproduces it poorly. The stellar-velocity field in the very
central part of the galaxy also suggests the possible presence of
a local maximum, but much closer to the center---at a
galactocentric distance of ${\simeq} 250$~pc~[34], which
corresponds to the circumnuclear disk.

A remarkable dynamical feature of this galaxy is that the observed
gas velocity (small circles in Fig.~5a) at $r>3$~kpc differs only
slightly from the rotational velocity of the stars (straight and slanted
crosses in Fig.~5a). In the case of an
axisymmetric, stationary disk, no model with a dynamically cool
gaseous disk can reproduce this behavior: for the usually
adopted gas velocity dispersion of 10~km/s, the stellar rotation
curve should be lower than the gas rotation curve by
20--30~km/s. The SAURON-VII~data~[17] also confirm the high
rotational velocity of the stellar component (the diamond and
triangle corresponding to the SAURON archive data in Fig.~5a).

The observed similarity of the rotational velocities of the old stars
and gas can, in principle, be explained if the turbulent
velocity of the gaseous medium is close to the stellar
velocity dispersion (30--70~km/s). In this case, the HI forms a thick
disk that is roughly the same as the disk of old stars. However, the
origin of such a dynamical peculiarity of the atomic-gas layer is by
no means evident. The galaxy lacks a sufficient number of young
stars to impart the required energy to the gas. The similarity in
the old-star and gas rotational velocities is more likely due to the
complex internal structure observed in the central part of the galaxy (a
high-contrast bar and short spiral arms), which may be
responsible for non-circular gas motions. In this case, the
rotational velocity of the stellar disk should be preferred over
that of the gaseous disk when constructing a model for the galaxy.

Figure~5b compares the observed line-of-sight stellar velocity
dispersions $c_{\textrm{obs}}$ and the velocity
dispersions~$c_\ell$ obtained for the marginally stable disk
model. In this model, the component masses within $r=12$~kpc are
$M_h/M_d=0.5$, $M_b/M_d=0.13$, and $M_d=8.05\times
10^{10}$~$M_{\odot}$. The estimated velocity dispersion
corresponding to the stability limit decreases with decreasing
disk mass fraction of the model. The significant difference
between $c_\ell$ and $c_{\textrm{obs}}$ in the central region of
the galaxy (up to 40~km/s) appears to be due to stars of the
dynamically ``hotter'' bulge; i.e., it does not refer to the disk.
At galactocentric distances $r=2{-} 5$~kpc, the observed velocity
dispersion is close to the expected velocity dispersion for this
model. The disk component may be slightly ``overheated'' at the
periphery of the stellar system ($r=7{-} 11$~kpc), but we must
bear in mind the large scatter of the estimated velocity
dispersions. We can thus conclude that a model with a marginally
stable disk having a close-to-maximum mass is consistent with the
observational data for this galaxy.

Note that the adopted parameters for the model with a marginally
stable disk can reproduce both the rotation curve and the
formation of a bar in the galaxy. In numerical simulations, a bar forms in
the inner part of the disk during one to two rotational periods as
the disk approaches the quasi-stationary state, as a result of
the disk's instability against the bar-forming mode.
However, the observed spiral structure (fragments of thin rings)
cannot be reproduced in collisionless models; a cool component
(gas) is evidently required for such structure to form.

The average vertical disk scale height in \linebreak
NGC~2273
calculated in the maximum-disk model is 0.8~kpc, which is twice
this parameter in the Milky Way Galaxy.

\textbf{NGC~4150.} Figure~6a illustrates the radial distributions
of the rotational velocity of the stellar disk (various symbols),
the circular rotational velocity calculated for the maximum-disk
model (solid bold curve), and the results of
decomposing the circular velocity for this model (thin solid
curves). The dynamical model of this galaxy was calculated with a
``live'' bulge, where the distributions of mass and particle
velocities could evolve with time.

In the maximum-disk model, the decomposition of the rotation curve
yields a central disk surface density of  $\sigma_0\approx 1330
$~$M_{\odot}/$pc$^2$ and a component-mass ratio of $M_h/M_d=0.6$.
The bulge mass fraction in this galaxy is low, $M_b/M_d=0.09$, but
the observed stellar kinematics cannot be explained by allowing
for the bulge.

Figure~6b shows the velocity dispersion estimates along the major
and minor axes for the galaxy model considered together with
the observational data. Again, the stellar disk is appreciably
overheated. The disk of the galaxy is fairly thick,
with a vertical-to-radial scale ratio of no less than 0.32. The
velocity dispersion of the ``live'' bulge in the model agrees
with the observed velocity dispersion at $r<0.8 $~kpc.

The photometric data imply a very short (less than 1~kpc) radial
disk scale for this galaxy. Given the slow radial decrease of the
stellar velocity dispersion, this implies that the thickness of
the stellar disk increases significantly with galactocentric
distance, so that the $z_0$ estimate listed in Table~3 below
should not be taken too literally. The disk half-thickness within
1~kpc from the center of this galaxy does not exceed 400~pc (which
is still greater than the corresponding parameter for the Milky
Way), but it becomes equal to the radial disk scale at a
galactocentric distance of several kpc. Starting from $r\approx
4{-} 5$~kpc, the disk proper appears to be absent. Indeed, the
photometric profile becomes flat at these galactocentric
distances~[12].

\begin{figure}
%%% Figure:6
\includegraphics[scale=0.9]{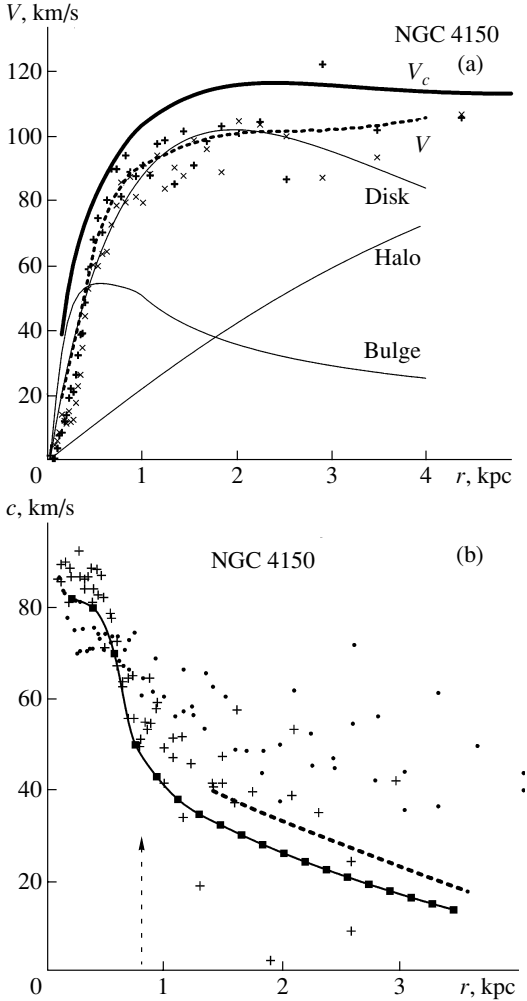}
\caption{NGC~4150. Shown are the (a) rotational velocity of the
stellar disk according to our measurements (straight and slanted
crosses show the measurements made along the major axis on either
side of the center), circular velocity in the adopted model (solid
bold curve), the components of the rotation curve corresponding to
the disk, bulge, and halo (thin solid curves), and the rotational
velocity of the thick disk in the maximum-disk model (dotted curve);
(b) stellar velocity dispersion $c_{\textrm{obs}}$ according to our
measurements along the major axis (crosses) and minor axis (small
filled circles), and line-of-sight velocity dispersion of disk stars
in the maximum-disk model along the major (large squares) and minor
(dotted curve) axes. The vertical dashed arrow indicates the
conventional boundary of the bulge.}
\end{figure}

\begin{figure}[t!]
%%% Figure:7
\includegraphics[scale=0.9]{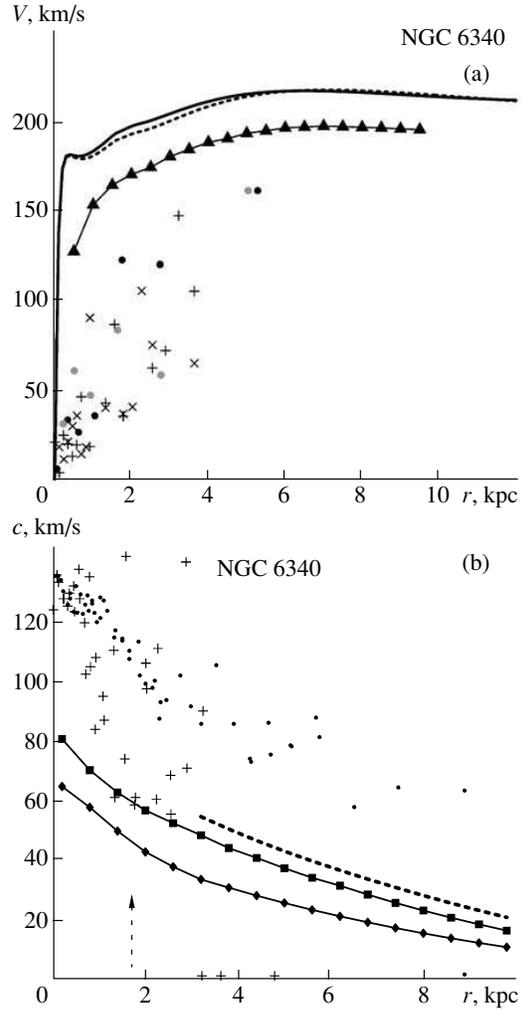}

\caption{NGC~6340. Shown are the (a) rotational velocity of the
stellar disk according to our measurements (straight and slanted
crosses show measurements made along the major axis on either side
of the center), the same for the data of Bottema~[35] (small filled
circles), the circular velocity for the maximum-disk model (bold
solid curve), and the rotational velocity of the stellar disk in
this model (filled triangles); (b) stellar velocity dispersion
$c_{\textrm{obs}}$ according to our measurements along the major
axis (crosses) and minor axis (small filled circles), the
line-of-sight velocity dispersion of the disk stars in the
maximum-disk model along the major (large squares) and minor (dotted
curve) axes, and the stellar velocity dispersion for the model with
central disk surface density $\sigma_0=1000 $~$M_{\odot}/$pc$^2$
(diamonds). The vertical dashed arrow indicates the conventional
boundary of the bulge.}
\end{figure}

\textbf{NGC~6340.} As in NGC~4150,
rotation was measured only for the stellar component. Only the
gradient of the rotational velocity along the major axis can be
confidently estimated (see Section~3). The estimates of the stellar
velocity dispersion reported both here and in the
earlier paper of Bottema~[35] are characterized by a large
scatter. The dispersion at the center of the galaxy is
130~km/s, but this refers to the bulge, not the disk.
The velocity dispersion decreases considerably with increasing $r$.
Judging from the photometric profile of the galaxy~[13], the disk
dominates in brightness beginning from $r\approx 1.7$~kpc.

Figure~7a shows the model rotation curves and measured rotational
velocities of the stellar disk. As we pointed out above, the
stellar-velocity measurements for this galaxy prevent the
reconstruction of the form of the rotation curve. Due to the lack
of direct estimates of the rotational velocity of the gaseous
component for this galaxy, we accepted  an overall form of the
rotation curve for the disk of this galaxy that makes it
consistent with the adopted radial scale $r_d$. The maximum of the
curve was conventionally set to be equl to  the HI rotational
velocity according to the HYPERLEDA
database---$V_{\textrm{max}}=219$~km/s (after reducing the latter
to the adopted inclination $i=26^{\circ}$ inferred from the outer
isophotes).

Figure~7b shows the radial dependences of the line-of-sight
velocity dispersion. At galactocentric distances 2--8~kpc, the
average estimated velocity dispersion based on measurements made
along the minor axis is no less than 60~km/s, suggesting that the
disk is rather ``hot.'' In the maximum-disk model, the central
disk surface brightness is $\sigma_0\approx 1380
$~$M_{\odot}/$pc$^2$. The velocity dispersion in the marginally
stable disk model is much lower than $c_{\textrm{obs}}$,
confirming the overheated state of the disk component. The
velocity-dispersion estimates extend beyond three radial scales
$r_d$ along the minor axis, far beyond the bulge. These estimates
suggest an approximately twofold excess of the observed stellar
velocity dispersion at the far edge of the disk over the minimum
required to maintain gravitational stability in the maximum-disk
model. Other evidence suggesting that the stellar system is
overheated includes the large systematic difference between the
rotational velocities of the stellar disk inferred from
observations and implied by the marginally stable disk model
(Fig.~7a).

\begin{table*}[t!]
%%% Table:3

\caption{Parameters of galaxy models. Note: The component masses
$M_d$, $M_b$, and $M_h$ refer to the region $r\leq
R_{\textrm{max}}$. Here, $z_0$ is the scale height of the vertical
density profile $\textrm{cosh}^{-2}(z/z_0)$.} \
\begin{tabular}{ccccccccc}
\hline \multicolumn{1}{c}{Model} & \multicolumn{1}{c}{$M_d$,}
&\multicolumn{1}{c}{$M_b$,} &\multicolumn{1}{c}{$M_h$,} &
$\sigma_0$, &\multicolumn{1}{c}{${M_h}/(M_d+M_b)$} &
\multicolumn{1}{c}{${M_d}/{L_B}$},&
\multicolumn{1}{c}{$z_{0}$,}&    \multicolumn{1}{c}{$z_0/r_d$}\\
& \multicolumn{1}{c}{$10^{10}$~$M_{\odot}$}&
\multicolumn{1}{c}{$10^{10}$~$M_{\odot}$}&
\multicolumn{1}{c}{$10^{10}$~$M_{\odot}$}& $M_{\odot}/$pc$^2$ &&${M_{\odot}}/{L_{\odot}}$&~kpc&     \\
 \hline
  \hline
  n1167-{a} & 38.7 & 12.0 & 26.0 & 1400 & 0.51 & 3.9 & 2.8\, & 0.35 \\
  n2273-{a} & 8.69 & 1.08 & 6.38 & 1100  & 0.65 & 5.9 & 0.8\, & 0.23 \\
  n4150-{a} & 0.53   & 0.046& 0.32 & 1330 & 0.56 & 1.6 & 0.78 & 0.93 \\
  n6340-{a} & 4.54   & 0.84 & 3.53 & 1380 & 0.66 & 3.8 & 1.3\, & 0.54\\
\hline
\end{tabular}
\end{table*}

\section{DISCUSSION AND CONCLUSIONS}

\textbf{Dynamical state of the disk.} Of the four galaxies
considered, the observed parameters of only one---the SBa galaxy
NGC~2273---are consistent with the hypothesis that the stellar
disk was not subject to dynamical heating, with the stellar
velocity dispersion corresponding to the minimum level sufficient
to maintain a quasi-stationary equilibrium state. In this respect,
NGC~2273 resembles many later-type spiral galaxies, whose stellar
velocity dispersions are close to their threshold levels
throughout a considerable fraction of the disk (see, e.g., [8, 22,
36, 37]).

The disks in the other three (lenticular) galaxies are ``hotter''
even for the maximum allowed mass. This can be viewed as evidence
that these galaxies were subject to external gravitational
perturbations in the past, most likely as a result of close
interactions with nearby galaxies or mergers of fairly large
satellites. Numerical simulations (see, e.g., [38]) confirm the
efficiency of the latter process. Gravitational perturbations of
the gaseous component of the disk can also explain why these
galaxies have rapidly exhausted their interstellar-gas reserves
due to a burst of star formation, thereby acquiring the properties
characteristic of lenticular systems.

\textbf{Estimates of ${c_r}$, ${c_z}$ and the thickness of the
stellar disk.} In a disk that is stable against bending
perturbations, perturbations in its plane due, e.g., to stochastic
spiral arms, would primarily increase $c_r$, and hence decrease $c_z
/c_r$. Merging of satellites and their passage across the disk
should result in a more isotropic distribution of the velocity
dispersion. In this case, heating occurs mostly via bending
perturbations of the disk (see Ardi et~al. [38] for the discussion
of this issue).

For all four galaxies, the velocity-dispersion components $c_z$ and
$c_r$ and their ratio $c_z/c_r$ calculated in the marginal-disk
model (Fig.~8) decrease with galactocentric distance. Note that
$c_z$ does not decrease rapidly enough to keep the marginally stable
disk thickness constant: it increases with $r$ in all cases. The
same is true for the measured stellar velocity dispersions in the
galaxies, although the insufficient accuracy of the
velocity-dispersion estimates prevented us from reconstructing the
detailed behavior of the radial variation of $c_z/c_r$ from the
observations. However, we estimated the average $c_z /c_r$ values by
analyzing the variation of the velocity dispersion along the major
and minor axes for three of the four galaxies observed---NGC~1167,
NGC~2273, and NGC~4150. We found this ratio to be 0.6--0.8 in the
interval (1--2)$r_d$ for all three galaxies, and to decrease to
0.4--0.5 at (2.5--3)$r_d$ (for NGC~2273 and NGC~4150).

\begin{figure}
%%% Figure:8
\includegraphics[scale=0.9]{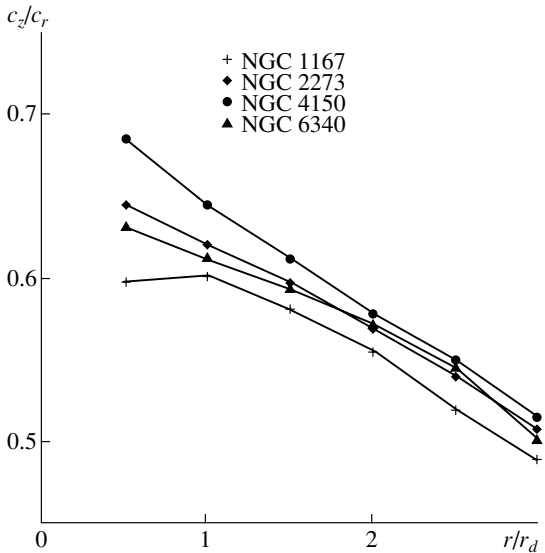}
\caption{Radial distributions of $c_z/c_r$ in numerical simulations
for maximum-disk models at the gravitational stability limit. }
\end{figure}

Table~3 lists the average disk half-thicknesses $z_0$ within the
maximum galactocentric radius covered by the observations for the
galaxies considered. We calculated these half-thicknesses for the
inferred  $c_z$ and vertical density profile
$\textrm{cosh}^{-2}(z/z_0)$. In all cases, the disks turned out to
be thicker than the disk of the Milky Way. The average $z_0$
estimate for NGC~4150 should not be taken at face value, it reduces
to a  half the tabulated value in the inner region of the galaxy
[(1--1.5)$r_d$] (see previous Section). The disk of this galaxy is
very ``chubby,'' strongly widens with $r$, so that it grades into
the halo at a distance of several radial disk scales.

Note that the inferred $z_0$ values are fairly typical for  the
thick disks of spiral  galaxies, where they coexist with more
massive thin disks of galaxies like those observed in the Hubble
Space Telescope Ultra Deep Field in early stages of their
formation~[39].

\textbf{Estimates of the masses of the disks and halos.} Table~3
lists estimates of the masses of these two main components of the
disk galaxies and the mass-to-luminosity ratio for the stellar
disk found for  the maximum-disk model. It is evident from the
estimates obtained that the mass of the dark halo is fairly high,
even when calculated in the maximum-disk model: the ratio of the
halo mass to the total mass of the stellar components
(disk~$+$~bulge) within the photometric radius is equal to
0.5--0.8. Similar or even higher ratios are also observed in
later-type galaxies~[29, 40]. Thus, lenticular galaxies do not
differ strongly from galaxies of other types in terms of their
dark-mass content.

The mass-to-luminosity ratio $M_d/L_B$ for the disks of the four
galaxies are several solar units, which, at least for three of these
objects (all except NGC~4150), is quite consistent with the ratio
expected for the old stellar population. For NGC~4150,
$M_d/L_B\approx 1.2$ in solar units. Although this ratio was
inferred for a model with the maximum disk mass, it is a factor of
two to three lower than expected for the othergalaxies. It is
unclear why the $M_d/L_B$ ratio is so low; such values are rarely
found in galaxies dominated by their old stellar population. Either
the disk of this galaxy is deficient in low-mass stars (due to a
shallow initial mass function), or the adopted model underestimates
the disk mass, e.g., due to an overestimation of the inclination
$i$. However, the latter possibility is unlikely, since the
inclination is fairly large (56$^{\circ}$), if our initial
assumption that the structure of the disk is axisymmetric is
correct.

Recall that NGC~4150 ranks well below the other three galaxies in
terms of its luminosity, mass (by about an order of magnitude), and
size. The stellar disk of this galaxy appears to have a low
volume density. The formation history and evolution of the disk
in this case may have differed from those for the disks in the other
more massive galaxies.

\section*{ACKNOWLEDGMENTS}

This paper is based on observational data obtained with the 6-m
telescope of the Special Astrophysical Observatory of the Russian
Academy of Sciences, which is operated with financial support from
the Ministry of Science of the Russian Federation (registration
number 01-43). Some of the observations at the 6-m telescope were
carried out by A.N.~Burenkov and S.S.~Ka{\u\i}sin. We are grateful
to Edo Nordermeer for sharing his digitized galaxy velocity fields
for NGC~2273 and NGC~4150 and to Eric Emsellem for kindly
providing velocity fields and stellar velocity dispersion from the
SAURON archive. This work was supported by the Russian Foundation
for Basic Research (project nos. 07-02-00792 and 07-02-01204).

%\newpage

\hfill\emph{Translated by A. Dambis}

%\endinput
\end{document}